\documentclass[conference]{IEEEtran}

\usepackage{url}
\usepackage{graphicx}
\graphicspath{{figures/}{./}}
\usepackage{amsmath}
\usepackage{booktabs}
\usepackage{array}
\usepackage{microtype}
\usepackage{balance}
\usepackage[numbers,sort&compress]{natbib}
\usepackage{float}
\usepackage{placeins}
\usepackage{hyperref}
\usepackage{hyperref}
\hypersetup{
    colorlinks   = true,
    citecolor    = blue,
    urlcolor    =  blue
}
\usepackage{makecell}

\newcommand{\hf}[2]{\raisebox{-2.2pt}{\includegraphics[scale=0.025]{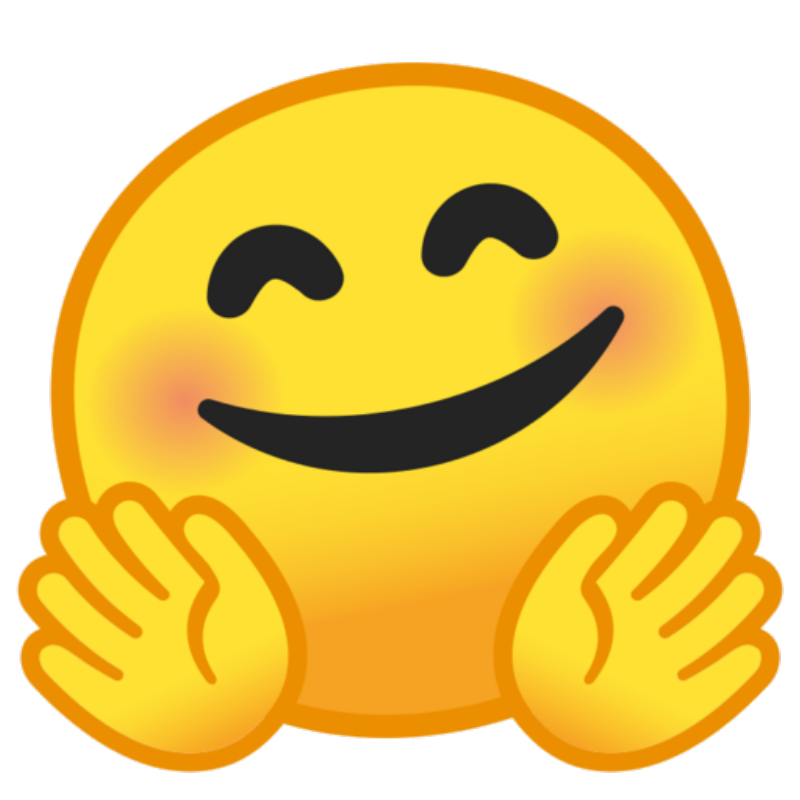}}~\href{#1}{\texttt{#2}}}

\newcommand{\gh}[2]{\raisebox{-2.2pt}{\includegraphics[scale=0.02]{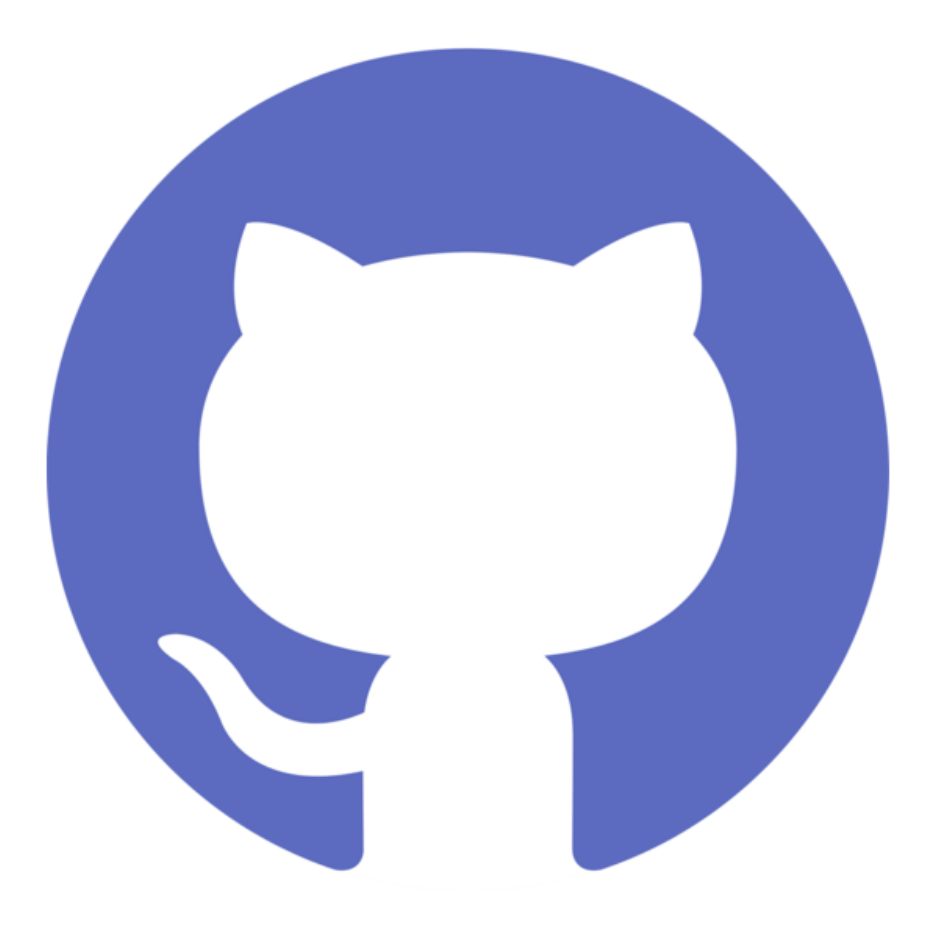}}~\href{#1}{\texttt{#2}}}

\begin{document}

\title{
Monsoon Mayhem to Market Waves:\\
Forecasting Fisheries Resilience in Sri Lanka
}

\author{
\IEEEauthorblockN{%
Ruzaini Ahmedh,
Yohan Jayasinghe,
Tharumini Gamage,
Ifaz Ikram,
Hasini Lawanya,\\
Nirasha Munasinghe,
Patalee Narasinghe,
Nisansa de Silva,
Sandareka Wickramanayake}
\IEEEauthorblockA{Dept.\ of Computer Science \& Engineering, University of Moratuwa, Sri Lanka.\\
\texttt{\{ruzainia.23, yohanj.23, tharuminig.23, ifazi.23, lawanyakkhg.23,}\\
\texttt{nirasha.25, patalee.21, NisansaDds, sandarekaw\}@cse.mrt.ac.lk}
}
}

\maketitle

\begin{abstract}
Sri Lanka's fisheries sector is important for jobs and food supply. Between 2019 and 2025, it faced several major problems at the same time, and how these events together affected fish production and prices is still not well understood.
This study develops a unified framework to connect weather changes, major disruption events, fish production, and prices, addressing the limitation of prior studies that analyse these factors in isolation. By integrating climate lag analysis, disruption estimation, and forecasting into a single pipeline, this framework provides actionable insights for policymakers, traders, and supply chain managers.
Seasonal patterns are studied using STL decomposition. Spearman lag correlation is used to find the delayed effects of climate on production. Interrupted Time Series (ITS) regression measures the impact of major events. SARIMAX models predict monthly production and prices. Hotspot detection identifies unusual patterns.
The results show that marine and inland fisheries behave differently in terms of seasons and climate effects. Our forecasting evaluation demonstrates that the proposed SARIMAX models reduce Root Mean Square Error (RMSE) by up to 66.0\% and maintain Mean Absolute Percentage Error (MAPE) between 4.71\% and 11.31\% compared to seasonal-na\"{i}ve baselines. Major disruptions caused different levels of impact; while inland production appeared to increase during periods of marine decline, suggesting a possible compensatory effect, this interpretation cannot be fully verified without fish import data.
These findings can support better planning, for example, improving infrastructure in high-risk areas, strengthening cold storage systems, and using early warning alerts for unusual events. Price forecasting tools should be used as decision-support tools, not as direct market signals.

\end{abstract}

\IEEEpeerreviewmaketitle

\section{Introduction}
Sri Lanka's fisheries sector supports rural employment and food security~\cite{MFARD2023Fisheries,SriLanka2022Food}. It covers four production types: offshore fleets, coastal fisheries, inland capture, and shrimp farming. Prices are observed weekly across national markets. Accurate price forecasts help traders decide when to sell, assist policymakers in planning interventions, and support supply chain coordination~\cite{lokanathan2016potential,lokanathan2014using} under volatile conditions.

Existing climate-fisheries studies focus on large industrial and temperate systems~\cite{cheung2010large,pinsky2013marine}. Sri Lanka-specific research is largely descriptive or limited to aggregate totals~\cite{pushpalatha2022climate,jayawardena2024climate}. Existing studies generally analyse climate impacts, disruption events, or fisheries forecasting in isolation. To our knowledge, no prior work integrates climate-lag analysis, interruption assessment, and forecasting within a unified framework. This combined analysis is essential because Sri Lanka's fisheries systems are simultaneously affected by environmental shifts and sudden socio-economic shocks. Treating these factors separately fails to capture the compound effects that determine actual supply constraints and price volatility.

Between 2019 and 2025, the sector faced overlapping shocks: the 2019 Easter attacks\footnote{\url{https://www.bbc.com/news/world-asia-48010697}}, COVID-19 disruptions\footnote{ \url{https://www.who.int/srilanka/emergencies/covid-19}}, the 2022 fuel and foreign exchange crisis~\cite{cbsl2023annual}, and 2024 floods\footnote{\url{https://reliefweb.int/disaster/fl-2024-000189-lka}}. Their combined impact on production and prices has not been measured at the category level.

This study addresses that gap by presenting a unified framework. Five objectives are pursued: (1)~characterise seasonal patterns and climate-production relationships; (2)~quantify price volatility; (3)~estimate disruption impacts via ITS; (4)~benchmark seasonal-na\"{i}ve and SARIMAX forecasts at monthly granularity, with same-week nowcasts at weekly granularity; and (5)~translate findings into resilience measures.

Our contribution combines climate-lag analysis, ITS-based disruption estimation, and category-level price forecasting in one integrated framework, keeping marine and inland systems and retail and wholesale channels separate at both temporal scales.
\hf{https://huggingface.co/datasets/sl-fisheries/sri-lanka-fisheries-resilience-data}{Data} and \gh{https://github.com/YohanJaya/sri-lanka-fisheries-ds-research}{code} for this work are publicly available.

\section{Related Work}
\subsection{Climate-Fisheries Linkages}
Foundational studies established that ocean warming shifts fish populations toward cooler regions, with tropical fisheries facing the largest projected declines~\cite{cheung2010large}, and that species track local climate velocities more closely than regional averages~\cite{pinsky2013marine}. However, most studies focus on large industrial fisheries in temperate and sub-Arctic regions, and do not directly apply to Sri Lanka's coastal and inland systems.

\subsection{Sri Lanka-Specific Studies}
Sri Lanka-specific studies are sparse and methodologically limited. \citet{pushpalatha2022climate} reviews climate risks across marine and inland sub-sectors without statistical modelling. \citet{jayawardena2024climate} document long-term marine production trends linked to monsoon patterns. \citet{dayaratne1995fish} provide an early account of small-scale fisher vulnerability to climate shocks in Sri Lankan lagoon systems, though without linking shocks to quantitative production or price data. Crucially, none of these Sri Lanka-specific studies employ predictive forecasting or robustly quantify specific historical disruption events on a category-level basis.

\subsection{Forecasting and Disruption Analysis}
Forecasting methods for fisheries prices commonly include AutoRegressive Integrated Moving Average (ARIMA) and seasonal ARIMA~\cite{box2015time,hyndman2018forecasting}, Prophet for additive decomposition-based forecasting~\cite{taylor2018forecasting}, and Random Forest for multivariate prediction~\cite{breiman2001random}. SARIMAX was selected over alternatives for this study because it natively handles explicit seasonal cycles alongside exogenous climate regressors, offering greater interpretability for policy-makers than opaque machine learning models like Random Forest, and better handling of strict ARIMA assumptions compared to Prophet. STL~\cite{cleveland1990stl} is widely used to identify seasonal patterns and support model selection.

\begin{figure*}[!htb]
\centering
\includegraphics[width=0.99\linewidth]{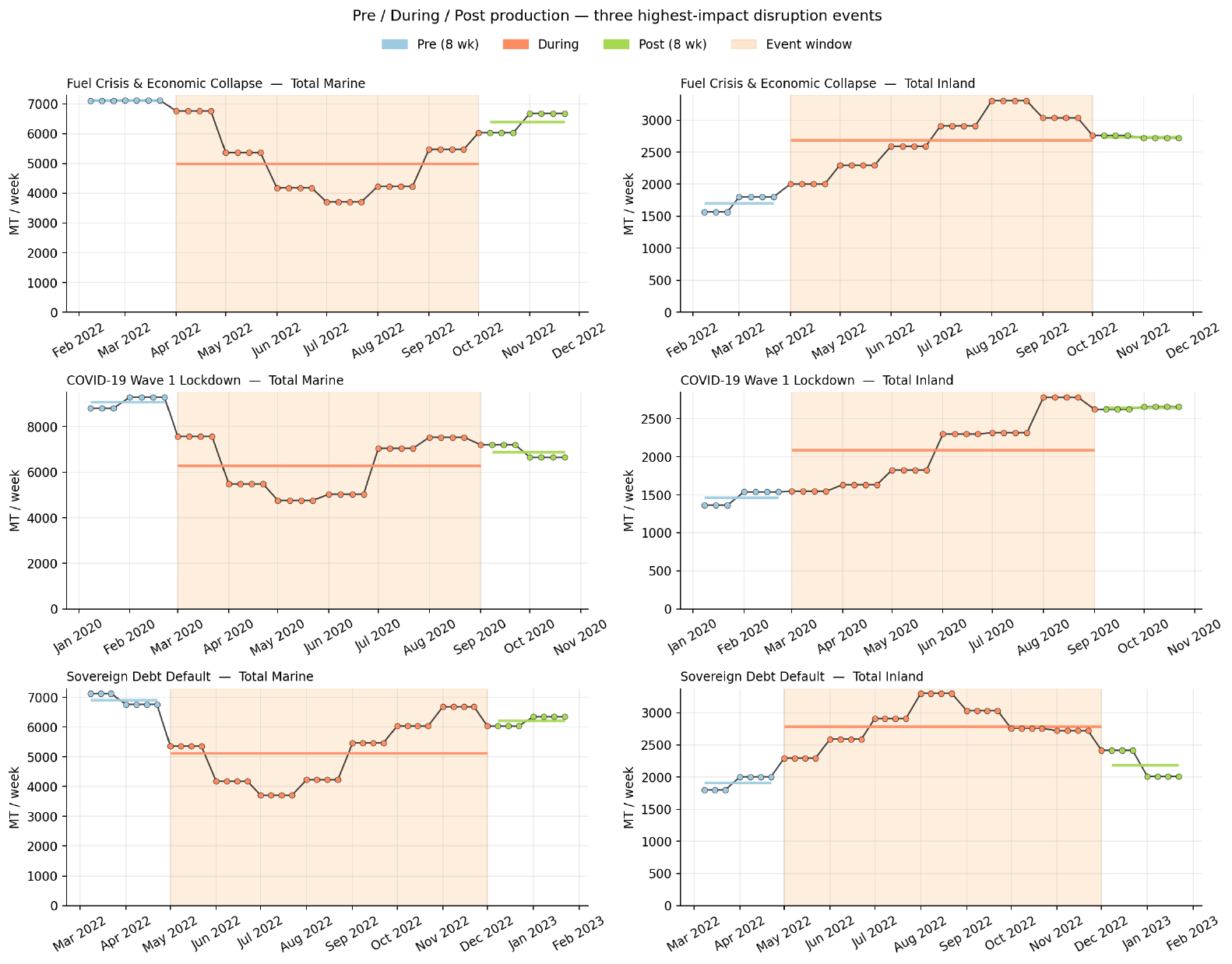}
\caption{Weekly fish production (Metric Tonnes, MT/week) in the 8-week pre-, during-, and post-event windows for the three highest-impact disruption events. Horizontal lines denote phase means; shaded regions mark event windows. These three events caused the largest production deviations across all 13 catalogued disruptions, motivating the ITS analysis.}
\label{fig:pre_during_post}
\end{figure*}

ITS regression is a standard quasi-experimental method for estimating whether a discrete event shifted the level or slope of a time series relative to the pre-event trend~\cite{bernal2017interrupted}. ITS is particularly suitable for these disruption events because it mathematically isolates the specific time windows of the shocks (e.g., lockdowns or floods) from the underlying seasonal trends, allowing for robust quantification of the immediate and prolonged effects of discrete historical crises. In Sri Lankan fisheries, the 2019 - 2025 period includes several major disruptions, but few studies have measured their effects by production category using ITS. Quantifying these effects has direct implications for food security policy, price stabilisation, and infrastructure investment in a sector that supports the livelihoods of over half a million people~\cite{MFARD2023Fisheries}. This study applies a segmented ITS design with pre-, during-, and post-event periods (Fig.~\ref{fig:pre_during_post}) to monthly marine and inland production and to category-level prices.

This gap is significant because Sri Lanka's fisheries underpin food security and rural livelihoods for millions of people~\cite{MFARD2023Fisheries,SriLanka2022Food}. An integrated, category-level analysis that combines climate effects, disruption impacts, and forecasting is therefore both scientifically novel and practically necessary.

\section{Data and Methodology}

\subsection{Data Sources}

Four data domains underpin this study. \textbf{Fisheries production:} Monthly data (2008--2025) from cumulative Department of Fisheries and Aquatic Resources(DFAR) terminal CSV reports~\cite{senaratna2025sri}, cross-verified for historical consistency. Production volumes (MT) are classified into \textit{Marine} (Offshore: deep-sea multi-day fleets; Coastal: near-shore artisanal fishing) and \textit{Inland} (Inland Capture, Aquaculture, Shrimp Farms). \textbf{Fish prices:} Weekly wholesale and retail prices for 30 species (2019--2025, $\approx$344 weeks) scraped from the DFAR portal; each record includes current, prior-week, and prior-year reference points exploited during imputation. Completeness is $\approx$91\% retail and $\approx$85\% wholesale. \textbf{Climate:} Six monthly variables retrieved from the NASA POWER Agroclimatology API (gap-free satellite record, 1981--present; MODIS, CERES, AIRS sensors), after ground-based records from the Department of Meteorology proved inaccessible despite formal requests. \textbf{Disruptions:} A catalogue of 13 events (2019--2024) including the Easter attacks, COVID-19 waves, and 2024 floods, each assigned timing, severity, and pre/during/post windows for ITS analysis.

\subsection{Bifurcation and Harmonisation}

The production dataset was bifurcated into Marine and Inland sub-datasets, as the drivers governing each domain (oceanic thermal gradients and monsoonal wind vs.\ reservoir recharge and ambient temperature) differ fundamentally. Production harmonisation involved three steps: (1)~aggregate totals re-derived programmatically from sub-sectors to correct arithmetic inconsistencies; (2)~non-alphanumeric artifacts stripped via regex; and (3)~the most recent entry prioritised where overlapping files conflicted. Price harmonisation involved four steps: (1)~vernacular name variants (e.g., \textit{Atawalla}/\textit{Kawakawa}) unified via a master nomenclature table; (2)~size-variant records aggregated to mean prices per species; (3)~inter-file conflicts resolved using the chronologically current entry; and (4)~missing values addressed by two-tier imputation-primary logic-based reconstruction using embedded ``week ago'' and ``year ago'' columns, with unresolvable gaps retained as missing to preserve seasonal variance. Exogenous shock gaps (Easter lockdown; COVID-19: retail May--November 2020, wholesale full-year 2020 to early 2021) were treated as genuine unavailability. In total, this harmonisation process resolved over 150 naming discrepancies and imputed approximately 4,200 missing price values, yielding a robust dataset of over 450,000 cleaned records. Prices were deflated to constant 2019 LKR using the World Bank CPI series.

\subsection{Meteorological Integration}

Six NASA POWER variables were used: T2M, T2M\_MAX, PRECTOTCORR, RH2M, WS2M, and WS10M. WS2M (surface turbulence, oxygen transfer) was assigned to inland analyses; WS10M (monsoonal energy, upwelling) to marine analyses. An 80-point spatial grid captures micro-climatic variability: the \textit{marine grid} (40 points) combines 20 coastal shoreline points at major lagoons and landing sites (Negombo, Batticaloa, Jaffna; 0-22~km) with 20 offshore shelf points (up to 160~km) across all four cardinal maritime zones; the \textit{inland grid} (40 points) covers reservoir cascades (Parakrama Samudra, Senanayake Samudra), river basins (Mahaweli, Kelani, Walawe, below 1,000~m), and aquaculture clusters. For each respective domain, the climate values were averaged equally across their 40 grid points without regional weighting to establish a unified national baseline. For the weekly price dataset, each of the 30 species is mapped to its ecological sub-sector and assigned the corresponding domain's meteorological variables row-level, preventing ecological misattribution in downstream models.

\section{Exploratory Data Analysis (EDA) and Climate Analysis}
\subsection{Seasonal Structure}

\begin{figure*}[!htb]
\centering
\includegraphics[width=0.9\linewidth]{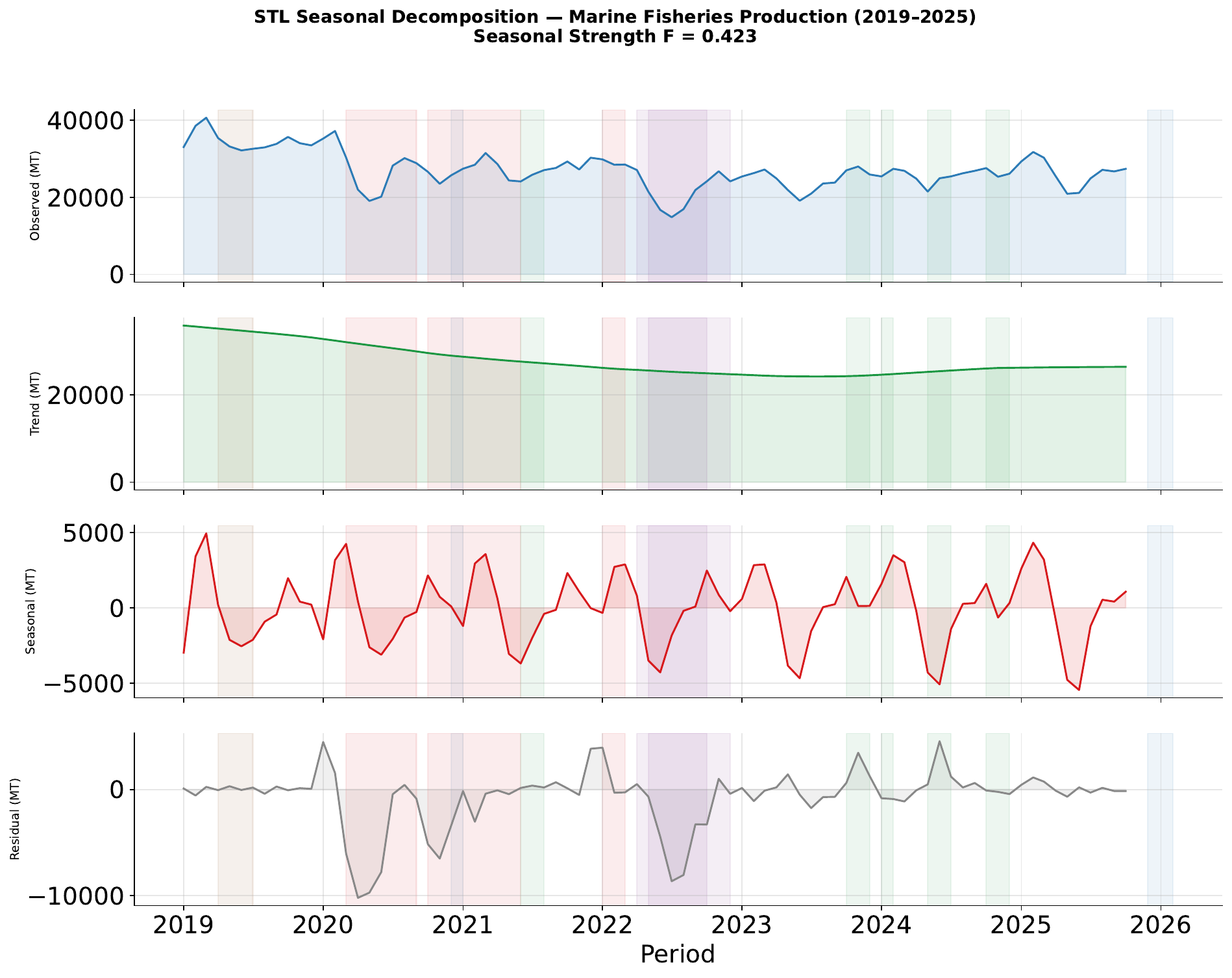}
\caption{STL decomposition of total marine production, showing trend, seasonal ($F_S=0.423$), and remainder components. The declining trend and strong monsoon-linked seasonal swings highlight operational vulnerability in marine fleets.}
\label{fig:stl_marine}
\end{figure*}

\begin{figure*}[!htb]
\centering
\includegraphics[width=0.9\linewidth]{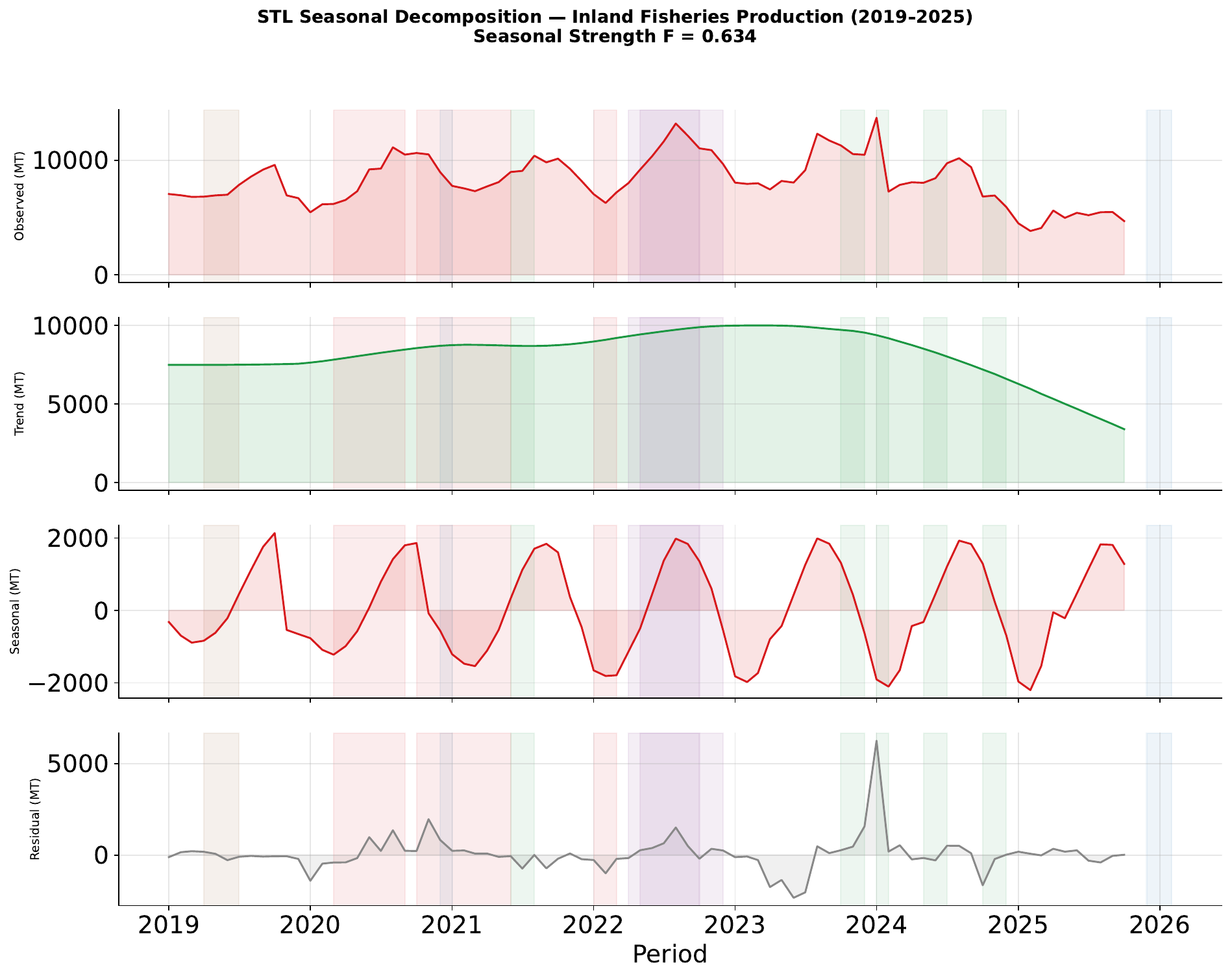}
\caption{STL decomposition of total inland production showing strong seasonality ($F_S=0.634$) with July-September peaks driven by monsoon-induced reservoir filling and flood-pulse dynamics. Stronger inland seasonality compared to marine systems motivates separate modelling.}
\label{fig:stl_inland}
\end{figure*}

STL~\cite{cleveland1990stl} separates each series into a \textit{trend}, a \textit{seasonal}, and a
\textit{remainder} component. The procedure was configured with a seasonal window of 13 months and a trend window of 21 months to capture stable annual cycles while remaining responsive to multi-year macroeconomic shocks. Seasonal strength $F_S \in [0,1]$ measures
how much of the non-trend variation follows a repeating annual pattern;
$F_S = 1$ denotes a perfectly regular cycle and $F_S = 0$ denotes no
seasonality~\cite{cleveland1990stl}.

Marine production (Fig.~\ref{fig:stl_marine}) shows moderate seasonality
($F_S = 0.423$), with a declining trend from $\sim$35{,}000~MT in 2019
to $\sim$23{,}000~MT by 2022 driven by COVID-19 and the fuel crisis.
Seasonal swings of $\pm$4{,}000~MT follow the monsoon cycle, yet large
irregular shocks in the remainder ($\text{Var}(R)/\text{Var}(S) = 1.365$)
suppress $F_S$ despite a consistent seasonal shape (year-on-year
$r = 0.982$).

Inland production (Fig.~\ref{fig:stl_inland}) exhibits stronger
seasonality ($F_S = 0.634$), with clear July-September peaks and
March-April troughs driven by rainfall-linked flood pulses and
aquaculture harvest cycles. Its remainder is considerably smaller
($\text{Var}(R)/\text{Var}(S) = 0.578$), reflecting the more controlled
nature of farm-based production. These structural differences motivate
separate modelling of the two systems.

\subsection{Climate-Production Relationships}

Lag-correlation analysis examines how prior-month climate variables relate to current production, capturing delayed biological and operational pathways. Lags of 0 - 3 months cover immediate weather effects (lag~0), within-season stock responses (lag~1 - 2), and slower biological cycles such as fish growth and aquaculture harvests (lag~3)~\cite{hyndman2018forecasting}. Spearman $\rho$ was used because Marine and Inland Rainfall failed the Shapiro-Wilk normality test ($p < 0.001$). A 3-month lag was included to capture delayed climate effects on fish growth and harvest scheduling~\cite{cheung2010large}. To rigorously account for the multiple testing framework across six climate variables and four lag periods, statistical significance was assessed at the $p < 0.05$ level, ensuring robust identification of primary climatic drivers.

\begin{figure*}[!htb]
\centering
\includegraphics[width=0.9\linewidth]{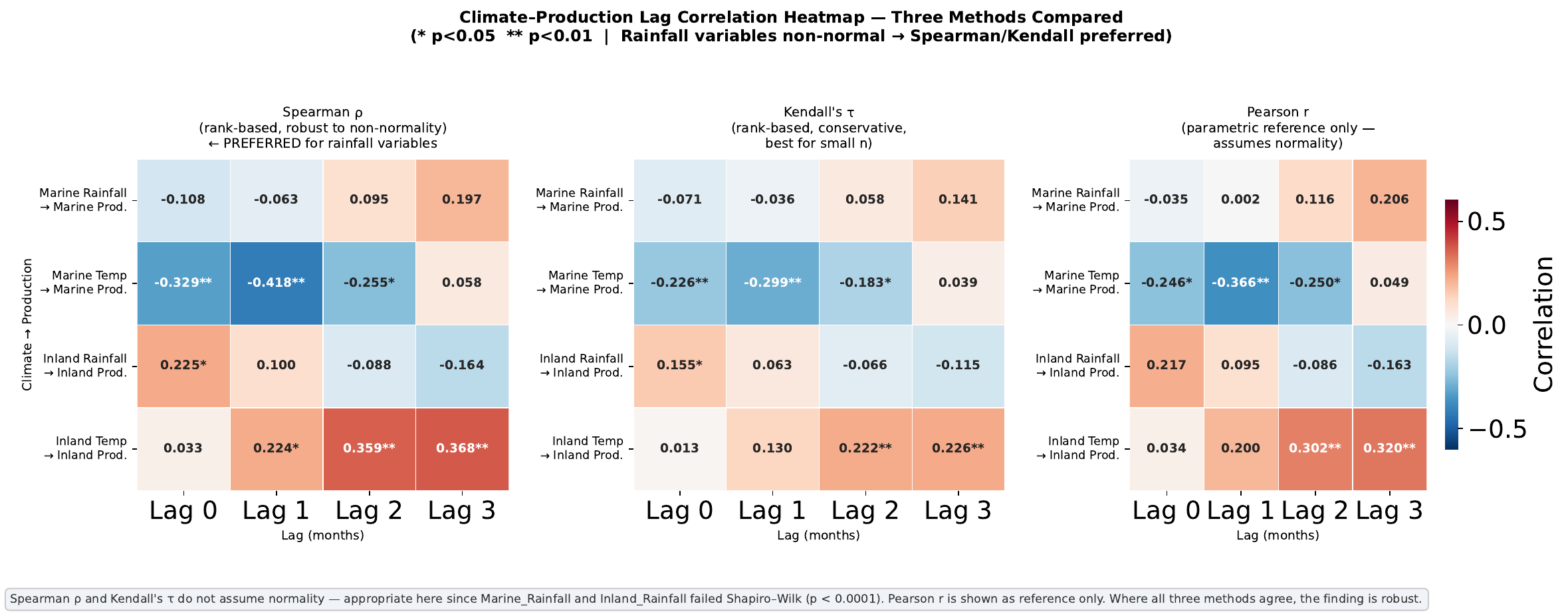}
\caption{Climate-production lag-correlation heatmap (Spearman $\rho$, lags 0-3 months). Stars denote statistical significance ($p<0.05$). Marine temperature suppresses production with a 1-month delay; inland temperature shows a positive effect growing through lag~3. These opposing lag structures form the empirical basis for including lagged climate regressors in SARIMAX.}
\label{fig:lag_heatmap}
\end{figure*}

Fig.~\ref{fig:lag_heatmap} reveals opposing climate responses between the
two systems. Marine temperature is the dominant driver ($\rho = -0.329$
at lag~0, peaking at $\rho = -0.418$ at lag~1, both $p < 0.01$),
indicating that warmer coastal waters suppress landings with a one-month
delay. Marine rainfall shows no significant relationship at any lag.
For inland systems, rainfall shows a modest same-month boost
($\rho = +0.225$, $p < 0.05$) that dissipates quickly, while inland
temperature strengthens progressively from lag~1 ($\rho = +0.224$) to
lag~3 ($\rho = +0.368$, $p < 0.01$), reflecting delayed growth and
productivity benefits of warmer preceding conditions. These contrasting
lag structures further support separate modelling of the two systems.

\subsection{Price Volatility and Disruption Catalogue}

\begin{figure*}[!htb]
\centering
\includegraphics[width=0.9\linewidth]{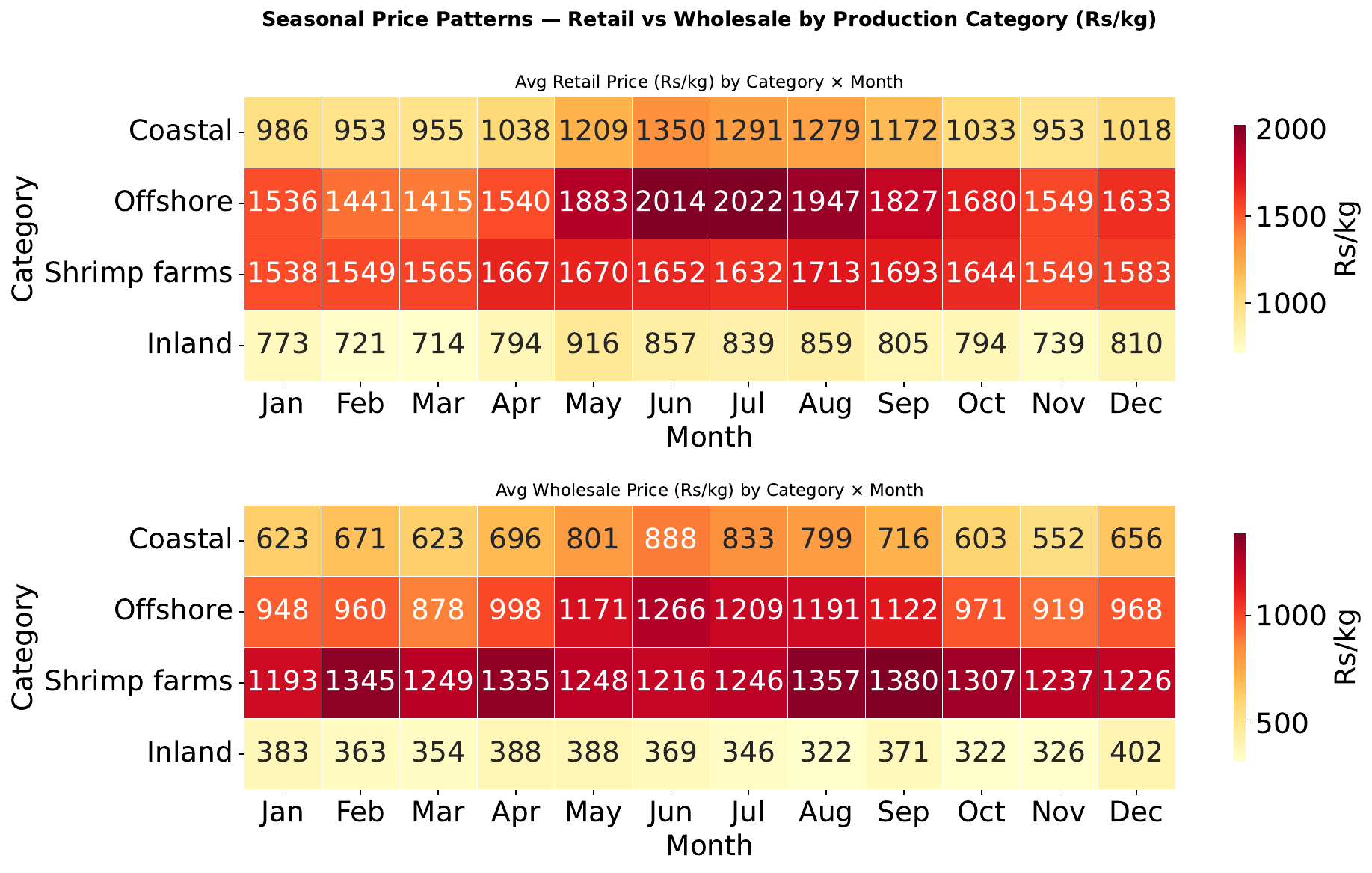}
\caption{Retail and wholesale price heatmaps by category and calendar month (constant 2019 LKR). Coastal and offshore categories show pronounced mid-year peaks, aligning with monsoon-period supply constraints that inform seasonal model specification.}
\label{fig:price_heatmap}
\end{figure*}

\begin{figure*}[!htb]
\centering
\includegraphics[width=0.9\linewidth]{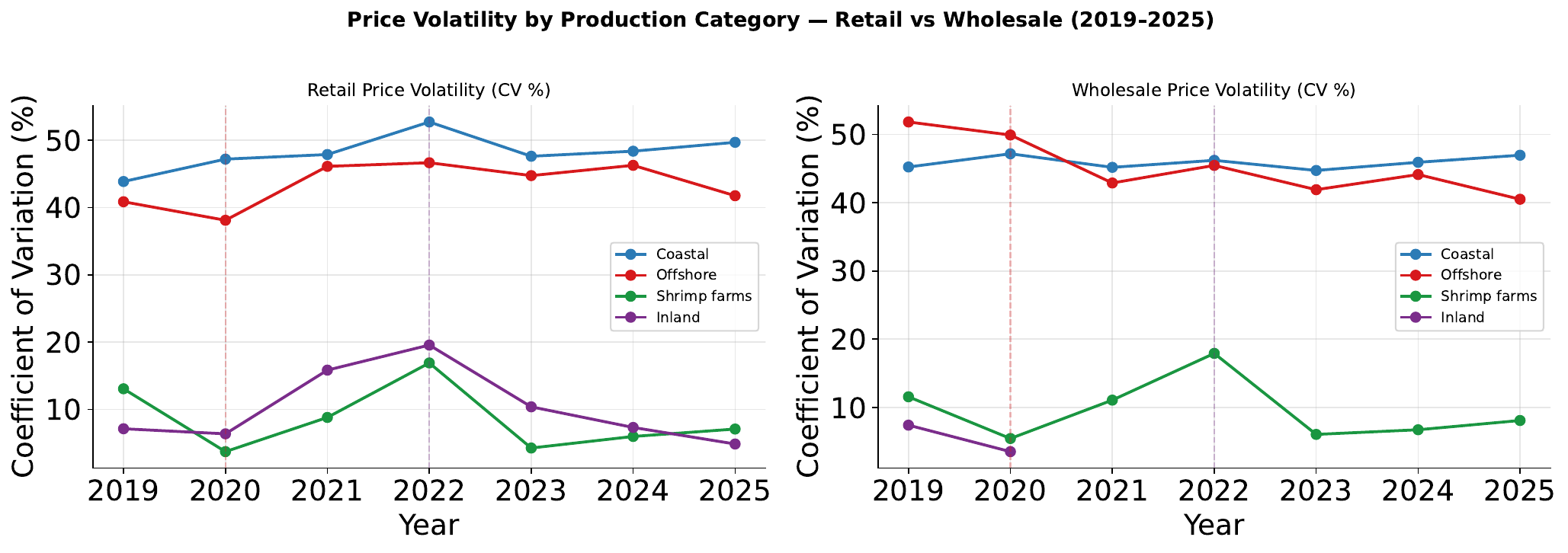}
\caption{Annual price volatility (CV) by production category and market channel (2019-2025). Retail CVs exceed wholesale CVs across all categories, identifying which channels most need targeted price stabilisation measures.}
\label{fig:vol_cv}
\end{figure*}

\begin{figure*}[!htb]
\centering
\includegraphics[width=0.9\linewidth]{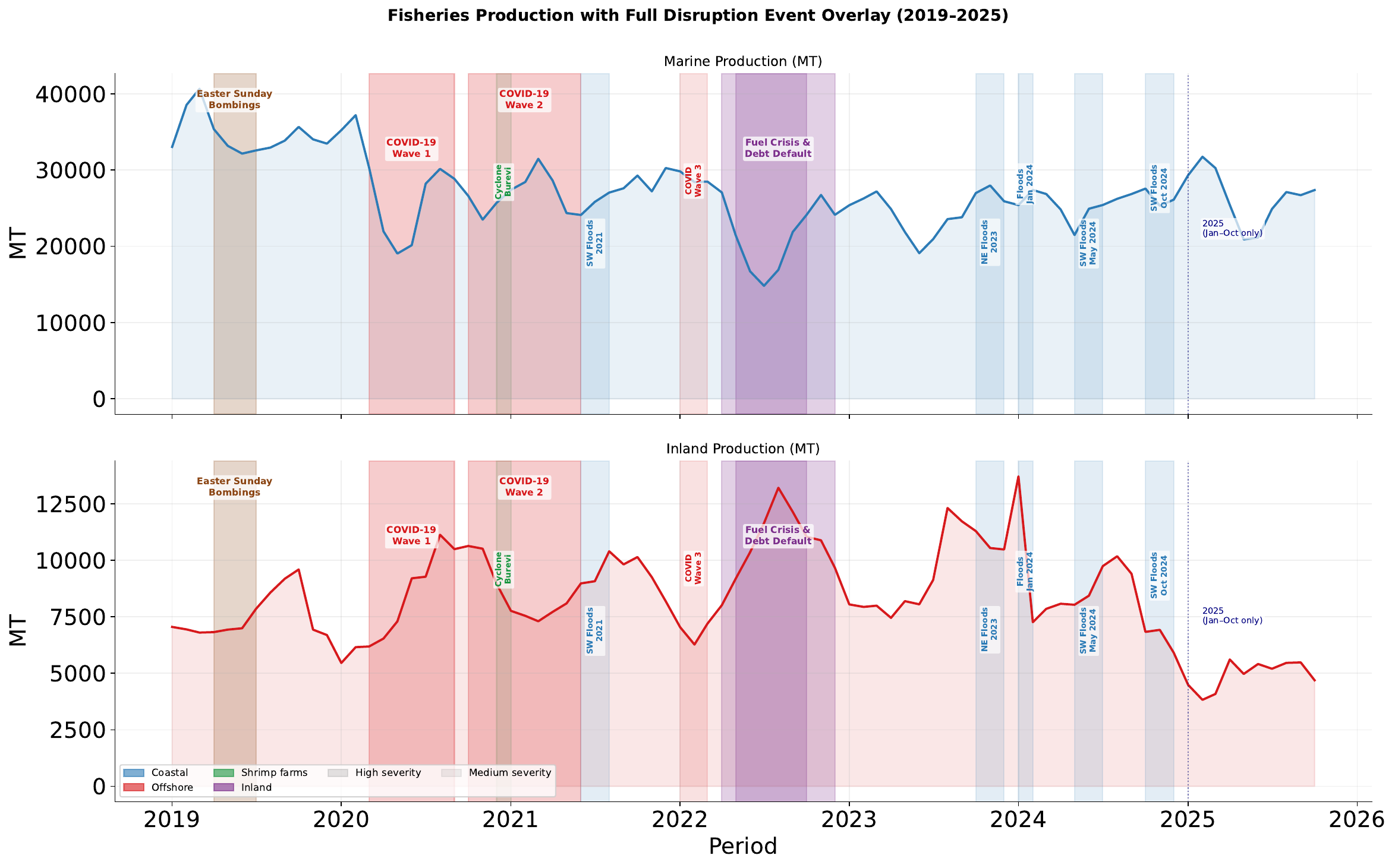}
\caption{Monthly marine production overlaid with major disruption events (2019-2025). Shaded regions indicate event windows; different shades represent varying severity. Marine output frequently contracts during major shocks while inland systems show partial substitution, motivating the ITS framework for category-level impact estimation.}
\label{fig:disruption_overlay}
\end{figure*}

Retail and wholesale price heatmaps (Fig.~\ref{fig:price_heatmap}) show clear mid-year peaks for coastal and offshore categories, while inland and shrimp farm prices remain more stable. Annual Coefficient of Variation (CV) analysis (Fig.~\ref{fig:vol_cv}) shows that retail prices vary more from year to year than wholesale prices across all categories, with the highest variation observed in 2022.

Disruption events are overlaid on monthly production (Fig.~\ref{fig:disruption_overlay}). The 2022 fuel and economic crisis~\cite{cbsl2023annual} led to a sharp decline in marine production, while inland production rose to its highest level during the study period. This inverse trend could indicate a partial substitution effect, where reduced marine supply shifted procurement toward inland and aquaculture species. However, fish import data are not available in this dataset, so this interpretation cannot be confirmed. During the 2022 currency crisis, cheaper imports may equally have displaced domestic marine demand rather than inland production filling a supply gap. Whether the marine decline represents a genuine domestic shortfall or a demand shift driven by import prices remains an open question that trade-level data would be needed to resolve.

\section{Modeling}
\subsection{Interrupted Time Series}
ITS regression estimates whether a discrete event shifted the level or slope of a time series relative to the pre-event trend~\cite{bernal2017interrupted}. ITS was applied to monthly national marine and inland production and to category-level retail and wholesale prices. Each model includes a linear time trend, sine/cosine seasonal terms, during- and post-event indicators, and a post-event slope term, estimated with Heteroskedasticity- and Autocorrelation-Consistent (HAC) standard errors. The ITS framework identifies associations with disruptions but does not establish full causal identification.

\subsection{Forecasting Architecture}
Forecasting was conducted using a walk-forward expanding-window framework, where each forecast uses only data available up to the prediction time, preventing information leakage.

The seasonal-na\"{i}ve model serves as the primary benchmark, using the same calendar month of the prior year for monthly series and the value from four weeks prior for weekly series~\cite{hyndman2018forecasting}. SARIMAX extends classical ARIMA with seasonal dependence and optional external covariates~\cite{box2015time,hyndman2018forecasting}. It was applied to monthly series: national marine production, national inland production, and category-level retail and wholesale prices. Model selection for each series was conducted via an exhaustive grid search minimising the Akaike Information Criterion (AIC), followed by walk-forward validation on an 80/20 train-test split to ensure robust out-of-sample generalisation. Candidate specifications varied by seasonal order (period 12), log transformation, training window (24 or 36 months), and optional lagged climate regressors at lags 1, 3, and 12 months. Retail price models used exogenous climate regressors; production and most wholesale models were selected as univariate specifications. Each model uses lagged target values, month-of-year sine/cosine encodings, a linear time trend, and lagged climate covariates where applicable.

Weekly price models are formulated as same-week nowcasts: rainfall and temperature are aggregated within the target week to estimate contemporaneous retail and wholesale prices by category. The weekly na\"{i}ve comparator uses the value from four weeks prior.

\subsection{Hotspot Detection and Ordinary Least Squares (OLS) Regression}
Hotspots are defined as the top 10\% of observations in each category-specific series to isolate extreme, economically destabilising price spikes while maintaining sufficient minority-class samples for training. Validation of the 10\% threshold confirmed optimal separation between routine seasonal variance and extreme disruption-linked events. A logistic regression model with class-weight adjustment to explicitly counter this imbalance is evaluated using F1-score (harmonic mean of precision and recall), Receiver Operating Characteristic Area Under Curve (ROC-AUC), and Precision-Recall Area Under Curve (PR-AUC). OLS models are fitted for monthly production totals using log-transformed outcomes, month fixed effects, lagged climate covariates, and a binary disruption indicator with robust Heteroskedasticity-Consistent (type HC3) standard errors.

\FloatBarrier
\section{Results}
\subsection{Regional Production Patterns}

\begin{table}[!htb]
\renewcommand{\arraystretch}{1.2}
\caption{Top-7 Inland Districts by Mean Monthly Production and CV (2014-2015)}
\label{tab:districts}
\centering
\begin{tabular}{lrr}
\toprule
\textbf{District} & \textbf{Mean (MT/mo)} & \textbf{CV (\%)} \\
\midrule
Anuradhapura & 1173.1 & 41.7 \\
Puttalam      &  799.6 & 57.8 \\
Monaragala    &  744.7 & 24.3 \\
Polonnaruwa   &  610.3 & 21.6 \\
Trincomalee   &  494.0 & 32.8 \\
Hambantota    &  439.6 & 30.9 \\
Ampara        &  338.3 & 59.6 \\
\bottomrule
\end{tabular}
\end{table}

District-level inland production data (2014 - 2015 only) are used for background context, as district-level data for 2019 - 2025 are unavailable. Despite the temporal mismatch, the 2014-2015 spatial variance fundamentally underpins the 2019-2025 national aggregate dynamics; understanding historical geographic vulnerabilities provides critical context for interpreting modern aggregate shock responses. Anuradhapura is the highest-producing district ($\sim$1,173~MT/month). Puttalam (CV~57.8\%) and Ampara (CV~59.6\%) combine high output with high volatility, while Monaragala (CV~24.3\%) and Polonnaruwa (CV~21.6\%) are more stable (Table~\ref{tab:districts}). These profiles highlight that national totals hide important variation, and district-level patterns remain relevant for infrastructure planning.

\subsection{Forecasting Performance}

\begin{table}[!htb]
\centering
\footnotesize
\caption{Forecasting performance comparison between seasonal-naïve and SARIMAX models}
\label{tab:forecast}
\resizebox{\columnwidth}{!}{%
\begin{tabular}{lrrrr}
\hline
Series & \makecell{S-Naïve\\RMSE} & \makecell{SARIMAX\\RMSE} & \makecell{SARIMAX\\MAPE (\%)} & \makecell{Improv.\\(\%)} \\
\hline
Inland Production (Total)      & 2545.08 & 1283.71 & 9.94  & 49.6 \\
Marine Production (Total)      & 4194.77 & 2156.42 & 7.28  & 48.6 \\
Coastal Price (Retail)         &  248.52 &  107.94 & 8.62  & 56.6 \\
Inland Price (Retail)          &  213.65 &   77.08 & 6.15  & 63.9 \\
Offshore Price (Retail)        &  284.58 &  135.65 & 7.92  & 52.3 \\
Shrimp Farms Price (Retail)    &  366.01 &  124.60 & 4.71  & 66.0 \\
Coastal Price (Wholesale)      &  208.05 &  106.50 & 11.31 & 48.8 \\
Offshore Price (Wholesale)     &  278.15 &  152.03 & 9.84  & 45.3 \\
Shrimp Farms Price (Wholesale) &  308.87 &  118.32 & 6.98  & 61.7 \\
\hline
\end{tabular}%
}
\end{table}

\begin{figure}[!htb]
\centering
\includegraphics[width=0.9\columnwidth]{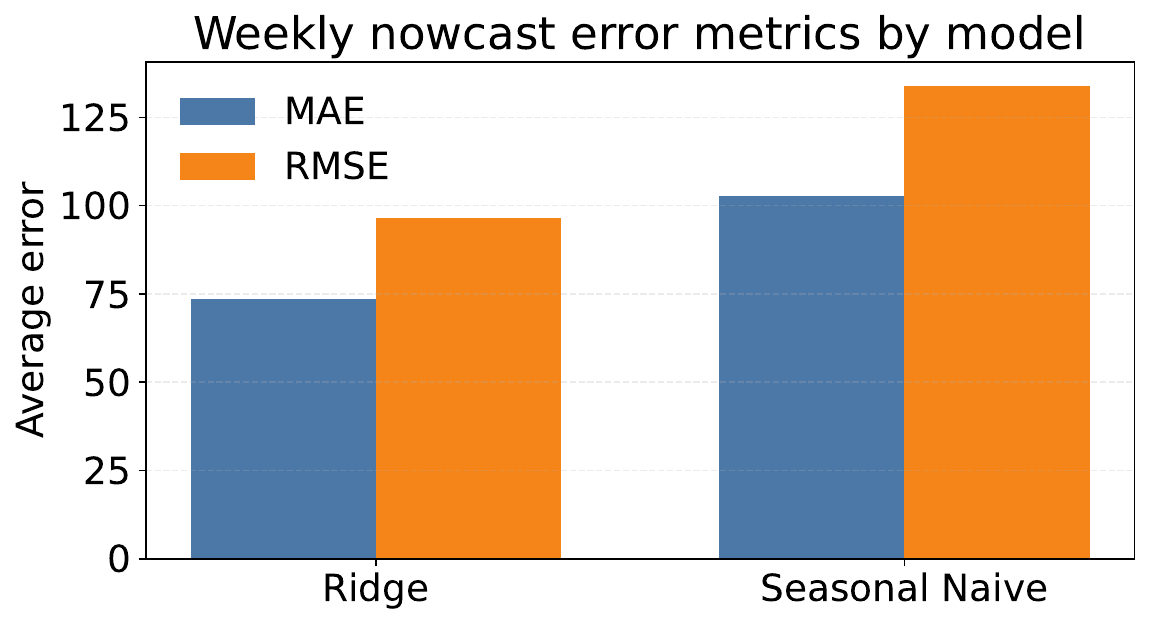}
\caption{Weekly forecast error metrics (MAE and RMSE) comparing SARIMAX and seasonal-na\"{i}ve baseline models. SARIMAX consistently produces lower errors across all price categories, justifying its use as the operational forecasting model.}
\label{fig:error_metrics}
\end{figure}

Walk-forward monthly forecasting demonstrates consistent improvements over seasonal-naïve baselines across both production and price series (Table~\ref{tab:forecast}, Fig.~\ref{fig:error_metrics}).

For inland production, SARIMAX reduces root mean square error (RMSE) by 49.6\% and achieves similarly substantial reductions in Mean Absolute Error (MAE) relative to the seasonal-na\"{i}ve baseline, with stable confidence intervals underscoring projection reliability. Marine production shows a 48.6\% RMSE improvement under SARIMAX. Retail and wholesale price series also show substantial gains, with RMSE improvements typically exceeding 50\% across major retail categories and ranging between approximately 45\% and 66\% across all evaluated series.

Mean Absolute Percentage Error (MAPE) ranged from 4.71\% to 11.31\% across all series, indicating strong predictive reliability.

\subsection{OLS Regression}
OLS models for monthly production totals show moderate fit (marine $R^2=0.367$, adj.~$R^2=0.217$; inland $R^2=0.409$, adj.~$R^2=0.269$). The inland disruption indicator is significant and positive (coeff.~$0.1859$, $p=0.0035$, 95\% CI [0.061, 0.311]), indicating an approximate 20\% production increase during disruption periods -- interpreted as a partial substitution effect when marine supply falls. A June fixed effect is significant for marine production (coeff.~$-0.4385$, $p=0.040$). The marine disruption indicator is not significant ($p=0.379$) in the aggregate OLS model; the ITS framework captures event-specific shifts more precisely by modelling each disruption with segmented trends (Table~\ref{tab:ols}). Although accuracy may be moderate in OLS, the directional findings remain practically useful for policy.

\begin{table}[!htb]
\renewcommand{\arraystretch}{1.15}
\caption{OLS: Key Coefficients for Log-Transformed Monthly Production (HC3 standard errors)}
\label{tab:ols}
\centering
\footnotesize
\resizebox{\columnwidth}{!}{%
\begin{tabular}{llrrl}
\toprule
\textbf{Model} & \textbf{Variable} & \textbf{Coeff.} & \textbf{$p$} & \textbf{95\% CI} \\
\midrule
Inland & Disruption        & $+0.186$ & 0.004 & [0.061,~~0.311] \\
Marine & June (fixed eff.) & $-0.439$ & 0.040 & [$-0.856$, $-0.021$] \\
Marine & Disruption        & $-$      & 0.379 & (n.s.) \\
\bottomrule
\end{tabular}%
}
\end{table}

\subsection{ITS Disruption Effects}
ITS analysis reveals substantial heterogeneity across events and sectors (Table~\ref{tab:its}). For marine production during COVID-19 Wave~1 (PAN01), the post-level effect is $-15{,}474$~MT ($p=0.006$) with a positive post-slope ($+517.97$, $p=0.033$). South-West (SW) Monsoon Floods 2021 (FLD01) produced a positive marine post-level ($+22{,}023$~MT, $p=0.007$) with a negative post-slope ($-660.9$, $p=0.027$). For inland production, the same 2021 flood event produced a negative post-level of $-8{,}034$~MT ($p=2.9\times10^{-6}$) with a positive post-slope ($+221.1$, $p=9.5\times10^{-4}$), while January 2024 floods (FLD03) caused a large positive post-level ($+13{,}279$~MT, $p{<}10^{-18}$) with a strongly negative post-slope ($-501.85$, $p{<}10^{-21}$).

\begin{table}[!htb]
\renewcommand{\arraystretch}{1.15}
\caption{Key ITS Post-Level and Post-Slope Effects by Event and System (HAC standard errors)}
\label{tab:its}
\centering
\scriptsize
\begin{tabular}{llrrrl}
\toprule
\textbf{Event} & \textbf{System} & \textbf{Level (MT)} & \textbf{$p$} & \textbf{Slope} & \textbf{$p$} \\
\midrule
PAN01 & Marine & $-15{,}474$ & 0.006 & $+517.97$ & 0.033 \\
FLD01 & Marine & $+22{,}023$ & 0.007 & $-660.90$ & 0.027 \\
FLD01 & Inland & $-8{,}034$ & $2.9\times10^{-6}$ & $+221.10$ & $9.5\times10^{-4}$ \\
FLD03 & Inland & $+13{,}279$ & ${<}10^{-18}$ & $-501.85$ & ${<}10^{-21}$ \\
\bottomrule
\vspace{0.8ex}
\end{tabular}
\begin{minipage}{\columnwidth}
\footnotesize PAN01 = COVID-19 Wave 1; FLD01 = SW Monsoon Floods 2021; FLD03 = Jan 2024 Floods.
\end{minipage}
\end{table}

\subsection{Hotspot Detection}
Retail categories show strong separability: coastal, inland, and offshore each reach ROC-AUC and PR-AUC of 1.00; shrimp farms reach ROC-AUC~0.90 and F1~0.571. These exceptionally high scores for some retail categories are largely driven by strong price clustering during distinct monsoon constraints, though they may also be inflated by the small sample sizes inherent to the top 10\% threshold and remaining class imbalances despite weighting adjustments, indicating a need for cautious operational application. Wholesale categories are moderate: coastal and offshore reach ROC-AUC~0.90, F1~0.667; shrimp farms reach F1~0.625. Marine production hotspot detection is usable (ROC-AUC~0.889, F1~0.50). Inland production is weak (F1~0.00) due to class imbalance; longer histories or revised thresholds are needed before operational use.

\section{Discussion}
The results point to three practical observations. SARIMAX forecasts were substantially more accurate than seasonal-na\"{i}ve baselines across all series: RMSE improved by 49.6\% for inland production, 48.6\% for marine production, and exceeded 50\% for most retail price categories, with MAPE between 4.71\% and 11.31\%. Even with a relatively short training window, these margins are large enough to be of practical value for procurement and planning decisions.

The two production systems behave quite differently. Marine production has a moderate seasonal pattern and a declining trend from 2019 onward; inland production follows a stronger and more regular annual cycle tied to monsoon and aquaculture harvest timing. Applying the same policy response or forecast specification across both systems would not be appropriate.

Disruption impacts also varied considerably across events. COVID-19 Wave~1 was followed by a gradual marine production recovery, while the 2024 inland floods caused a sharp initial rise and then a steep decline. Furthermore, the observed inverse relationship between marine decline and inland growth suggests a possible compensatory effect; however, this remains a tentative explanation that requires fish import data to fully confirm. These differences are relevant for designing targeted responses: recovery trajectories depend on the event type and the production system affected. The results support prioritising infrastructure investment in high-volatility districts, using ITS estimates to anticipate recovery timing, and deploying the forecasting models as decision-support tools rather than direct market signals.

\section{Limitations}
Several limitations should be noted. Climate variables were aggregated monthly over broad spatial grids, which may obscure localised effects. The OLS models assume linear lagged relationships and cannot capture interaction effects or structural breaks. Hotspot detection is sensitive to class imbalance, which particularly affects inland production and sparse wholesale strata. For some event--series combinations, ITS analysis could not be completed due to insufficient pre- or post-event observations. District-level causal analysis was not possible because district-wise price data are unavailable for the study period; climate variables were therefore aggregated at the national level. CPI deflation to constant 2019 LKR partially addresses real price comparisons but does not account for transport cost changes or the severity of the 2022 inflation episode. Finally, fish import data were not available, which means the observed inland production increase during the 2022 crisis cannot be clearly separated from a possible import-driven displacement of domestic marine supply. Consequently, the hypothesis of sector substitution must be interpreted with strict caution. Additionally, the sensitivity of the hotspot detection model to class imbalance limits its immediate operational deployment without further refinement or larger training samples.

\section{Conclusion}
This study analysed Sri Lanka's fisheries sector from 2019 to 2025 using a combined approach: climate-lag analysis, ITS-based disruption estimation, SARIMAX forecasting, and hotspot detection.

Forecast accuracy under SARIMAX was substantially better than seasonal-na\"{i}ve baselines for both production and price series. ITS analysis estimated event-specific level and slope shifts for major disruptions, which can help planners anticipate both the initial impact and the recovery trajectory. District volatility profiles from the 2014--2015 background data point to where infrastructure investment is most needed, even though updated district-level data are unavailable. Hotspot detection performed well for retail categories and is ready for operational use in those series; inland production hotspot detection requires longer data or revised thresholds before deployment.

Price forecasts can support sell-vs-store decisions at the producer level and help supply chain coordinators time procurement. They are best treated as one input to planning decisions, not as autonomous market signals.

\balance

{\footnotesize
\bibliographystyle{IEEEtranN}
\bibliography{references_arxiv}

@article{cheung2010large,
  title={Large-scale redistribution of maximum fisheries catch potential in the global ocean under climate change},
  author={Cheung, William WL and Lam, Vicky WY and Sarmiento, Jorge L and Kearney, Kelly and Watson, REG and Zeller, Dirk and Pauly, Daniel},
  journal={Global change biology},
  volume={16},
  number={1},
  pages={24--35},
  year={2010},
  publisher={Wiley Online Library}
}

@article{pinsky2013marine,
  title={Marine taxa track local climate velocities},
  author={Pinsky, Malin L and Worm, Boris and Fogarty, Michael J and Sarmiento, Jorge L and Levin, Simon A},
  journal={Science},
  volume={341},
  number={6151},
  pages={1239--1242},
  year={2013},
  publisher={American Association for the Advancement of Science}
}

@article{pushpalatha2022climate,
  title={Climate Change Impact on Inland Fisheries and Aquaculture-Sri Lanka},
  author={Pushpalatha, KBC},
  journal={Impact of Climate Change on Hydrological Cycle, Ecosystem, Fisheries and Food Security},
  pages={151--161},
  year={2022},
  publisher={CRC Press}
}

@incollection{jayawardena2024climate,
  title={Climate variability, observed climate trends, and future climate projections for Sri Lanka},
  author={Jayawardena, IM Shiromani Priyanthika and Darshika, DWTT and Herath, HMRC and Hapuarachchi, HASU},
  booktitle={The role of tropics in climate change},
  pages={77--119},
  year={2024},
  publisher={Elsevier}
}

@article{dayaratne1995fish,
  title={Fish resources and fisheries in a tropical lagoon system in Sri Lanka.},
  author={Dayaratne, P and Gunaratne, ABAK and Alwis, MM},
  year={1995},
  vol={24},
  no={7/8},
  pages={402-410},
  journal={Ambio}
}

@book{box2015time,
  title={Time series analysis: forecasting and control},
  author={Box, George EP and Jenkins, Gwilym M and Reinsel, Gregory C and Ljung, Greta M},
  year={2015},
  publisher={John Wiley \& Sons}
}

@book{hyndman2018forecasting,
  title={Forecasting: principles and practice},
  author={Hyndman, Rob J and Athanasopoulos, George},
  year={2018},
  publisher={OTexts}
}

@article{taylor2018forecasting,
  title={Forecasting at scale},
  author={Taylor, Sean J and Letham, Benjamin},
  journal={The American Statistician},
  volume={72},
  number={1},
  pages={37--45},
  year={2018},
  publisher={Taylor \& Francis}
}

@article{breiman2001random,
  title={Random forests},
  author={Breiman, Leo},
  journal={Machine learning},
  volume={45},
  number={1},
  pages={5--32},
  year={2001},
  publisher={Springer}
}

@article{cleveland1990stl,
  title={STL: A seasonal-trend decomposition},
  author={Cleveland, Robert B and Cleveland, William S and McRae, Jean E and Terpenning, Irma and others},
  journal={J. off. Stat},
  volume={6},
  number={1},
  pages={3--73},
  year={1990}
}

@article{bernal2017interrupted,
  title={Interrupted time series regression for the evaluation of public health interventions: a tutorial},
  author={Bernal, James Lopez and Cummins, Steven and Gasparrini, Antonio},
  journal={International journal of epidemiology},
  volume={46},
  number={1},
  pages={348--355},
  year={2017},
  publisher={Oxford University Press}
}

@article{senaratna2025sri,
  title={Sri Lanka Document Datasets: A Large-Scale, Multilingual Resource for Law, News, and Policy},
  author={Senaratna, Nuwan I},
  journal={arXiv preprint arXiv:2510.04124},
  year={2025}
}

@book{cbsl2023annual,
author = {{Central Bank of Sri Lanka}},
  title = {Annual Economic Review 2022},
year = {2023},
publisher={CBSL}
}

@article{SriLanka2022Food,
author = {{Food and Agriculture Organization of the United Nations}},
  title = {Fishery and Aquaculture Country Profiles. Sri Lanka},
   url = "https://www.fao.org/fishery/en/facp/lka",
  journal={Fisheries and Aquaculture},
year = {2022}
}

@book{MFARD2023Fisheries,
author = {{Ministry of Fisheries and Aquatic Resources Development}},
  title = {Fisheries Statistics 2023},
year = {2023},
publisher={MFARD}
}

@article{lokanathan2016potential,
  title={The Potential of Mobile Network Big Data as a Tool in {Colombo}'s Transportation and Urban Planning},
  author={Lokanathan, Sriganesh and Kreindler, Gabriel E and de Silva, N. H. Nisana and Miyauchi, Yuhei and Dhananjaya, Dedunu and Samarajiva, Rohan},
  journal={Information Technologies \& International Development},
  volume={12},
  number={2},
  pages={pp--63},
  year={2016},
  misc={https://itidjournal.org/index.php/itid/article/download/1506/1506-4213-1-PB.pdf,BIG,https://i1.rgstatic.net/publication/314475194_Using_Mobile_Network_Big_Data_for_Informing_Transportation_and_Urban_Planning_in_Colombo/links/59dcf407a6fdcce237e2068f/largepreview.png,https://goo.gl/xrdNbF}
}

@inproceedings{lokanathan2014using,
  title={Using Mobile Network Big Data for Informing Transportation and Urban Planning in {Colombo}},
  author={Lokanathan, S and de Silva, N and Kreindler, G and Miyauchi, Y and Dhananjaya, D},
  journal={Available at SSRN},
   month={November},
  year={2014},
  misc={https://goo.gl/S3x7rU,BIG,https://goo.gl/iY6aTr,https://goo.gl/mjLMMc}
}
}

\end{document}